\documentclass[aps,prc,preprint,footinbib]{revtex4}

\makeatletter
\let\@ORGREVTEXendnotemark\@endnotemark
\let\@ORGREVTEX@makefnmark@cite\@makefnmark@cite
\def\@endnotemark{\bgroup\@fileswfalse\@ORGREVTEXendnotemark\egroup}
\def\@makefnmark@cite{\bgroup\@fileswfalse\@ORGREVTEX@makefnmark@cite\egroup}
\makeatother

\usepackage{mathrsfs}
\usepackage{amsfonts}
\usepackage{amssymb}
\usepackage{amsmath}
\usepackage{bm}
\usepackage{bbm}
\usepackage{slashed}
\usepackage{rotating}
\usepackage{dcolumn}
\usepackage{multirow}
\usepackage{mciteplus}

\newcommand{\beq}{\begin{equation}}
\newcommand{\eeq}{\end{equation}}
\newcommand{\beqa}{\begin{eqnarray}}
\newcommand{\eeqa}{\end{eqnarray}}

\newcommand{\bvec}[1]{\ensuremath{\mathbf{#1}}}

\newcommand{\lag}{{\lambda_\gamma}}

\newcommand{\lN}{\lambda_N}

\newcommand{\half}{\ensuremath{\frac{1}{2}}}
\newcommand{\thalf}{\ensuremath{\frac{3}{2}}}

\newcommand{\shalf}{\ensuremath{{\scriptstyle\frac{1}{2}}}}
\newcommand{\sthalf}{\ensuremath{{\scriptstyle\frac{3}{2}}}}

\newcommand{\nn}{\ensuremath{N\!N}}

\newcommand{\gonn}{\ensuremath{g_{\omega N\!N}}}
\newcommand{\gonna}{\ensuremath{g^t_{\omega N\!N}}}

\newcommand{\Lonn}{\ensuremath{\Lambda_{\omega N\! N}}}
\newcommand{\Lonna}{\ensuremath{\Lambda^t_{\omega N\! N}}}

\newcommand{\qmag}{|\bvec{q}|}

\newcommand{\Gbar}{\overline{\Gamma}}

\newcommand{\vmbmb}{\ensuremath{v_{M'B',MB}}}

\newcommand{\tjlsmngn}{\ensuremath{t^{JT}_{L'S'M'N',\lag\lN T_{N,z}}}}

\newcommand{\vjlsmngn}{\ensuremath{v^{JT}_{L'S'M'N',\lag \lN T_{N,z}}}}
\newcommand{\vjlsmbgn}{\ensuremath{v^{JT}_{LSMB,\lag \lN T_{N,z}}}}
\newcommand{\tjlsmnmb}{\ensuremath{t^{JT}_{L'S'M'N',LSMB}}}
\newcommand{\Tjlsmbmb}{\ensuremath{T^{JT}_{LSMB,L'S'M'B'}}}
\newcommand{\tjlsmbmb}{\ensuremath{t^{JT}_{LSMB,L'S'M'B'}}}
\newcommand{\tjlsmnpn}{\ensuremath{t^{JT}_{L'S'M'N',\ell \pi N}}}
\newcommand{\tjlsmbpn}{\ensuremath{t^{JT}_{LSMB,\ell \pi N}}}
\newcommand{\vjlsmnpn}{\ensuremath{v^{JT}_{L'S'M'N',\ell \pi N}}}

\newcommand{\vjlsmbpn}{\ensuremath{v^{JT}_{LSMB,\ell \pi N}}}

\newcommand{\Tjlsmbgn}{\ensuremath{t^{R,JT}_{LSMB,\lag \lN T_{N,z}}}}
\newcommand{\Tfjlsmbgn}{\ensuremath{T^{JT}_{LSMB,\lag \lN T_{N,z}}}}

\newcommand{\Tjlsmbpn}{\ensuremath{t^{R,JT}_{LSMB,\ell \pi N}}}
\newcommand{\Gbjlsi}{\ensuremath{{\Gamma}^{JT}_{LSMB,N^*_i}}}
\newcommand{\Gbjlspi}{\ensuremath{{\Gamma}^{JT}_{L'S'M'B',N^*_i}}}
\newcommand{\Gjlsi}{\ensuremath{\overline{\Gamma}^{JT}_{LSMB,N^*_i}}}
\newcommand{\Gijls}{\ensuremath{\overline{\Gamma}^{JT}_{N^*_i,LSMB}}}
\newcommand{\Gbijls}{\ensuremath{{\Gamma}^{JT}_{N^*_i,LSMB}}}
\newcommand{\Gjpn}{\ensuremath{\overline{\Gamma}^{JT}_{N^*_j,\ell\pn}}}
\newcommand{\Gign}{\ensuremath{\overline{\Gamma}^{JT}_{N^*_i,\lag\lN T_{N,z}}}}
\newcommand{\Gbign}{\ensuremath{{\Gamma}^{JT}_{N^*_i,\lag\lN T_{N,z}}}}
\newcommand{\Gjlsj}{\ensuremath{\overline{\Gamma}^{JT}_{LSMB,N^*_j}}}
\newcommand{\Gjem}{\ensuremath{\overline{\Gamma}^{JT}_{N^*_j,\lag\lN T_{N,z}}}}
\newcommand{\Ljtlsmbn}{\ensuremath{\Lambda^{JT}_{N^*LSMB}}}
\newcommand{\Drij}{\ensuremath{\mathcal{D}^{-1}_{ij}}}
\newcommand{\Mbres}{\ensuremath{M^{(0)}_{N^*}}}
\newcommand{\Cjtnlsmb}{\ensuremath{C^{JT}_{N^*LSMB}}}
\newcommand{\Ljtnlsmb}{\ensuremath{\Lambda^{JT}_{N^*LSMB}}}
\newcommand{\knstar}{\ensuremath{k_{N^*}}}

\newcommand{\vmbgn}{\ensuremath{v_{MB,\gamma N}}}
\newcommand{\vpngn}{\ensuremath{v_{\pi N,\gamma N}}}
\newcommand{\vongn}{\ensuremath{v_{\omega N,\gamma N}}}
\newcommand{\vonpn}{\ensuremath{v_{\omega N,\pi N}}}
\newcommand{\vpnpn}{\ensuremath{v_{\pi N,\pi N}}}
\newcommand{\vonon}{\ensuremath{v_{\omega N,\omega N}}}

\newcommand{\vonen}{\ensuremath{v_{\omega N,\eta N}}}
\newcommand{\vonpd}{\ensuremath{v_{\omega N,\pi\Delta}}}
\newcommand{\vonsn}{\ensuremath{v_{\omega N,\sigma N}}}
\newcommand{\vonrn}{\ensuremath{v_{\omega N,\rho N}}}

\newcommand{\gnpn}{\ensuremath{\gamma N\to \pi N}}
\newcommand{\gnmb}{\ensuremath{\gamma N\to MB}}
\newcommand{\gpop}{\ensuremath{\gamma p\to \omega p}}
\newcommand{\pnmb}{\ensuremath{\pi N\to MB}}
\newcommand{\pnon}{\ensuremath{\pi N\to \omega N}}
\newcommand{\gppzp}{\ensuremath{\gamma p\to \pi^0 p}}
\newcommand{\gpppn}{\ensuremath{\gamma p\to \pi^+ n}}
\newcommand{\onon}{\ensuremath{\omega N\to \omega N}}
\newcommand{\pmpon}{\ensuremath{\pi^- p\to \omega n}}
\newcommand{\pnpn}{\ensuremath{\pi N\to \pi N}}

\newcommand{\Gmb}{\ensuremath{G_{0,MB}}}

\newcommand{\eps}{\ensuremath{\epsilon}}

\newcommand{\komg}{\ensuremath{\kappa_\omega}}
\newcommand{\komga}{\ensuremath{\kappa^t_\omega}}

\newcommand{\dspdo}{\ensuremath{{\frac{d\sigma_\pi}{d\Omega}}}}
\newcommand{\dsgdo}{\ensuremath{{\frac{d\sigma_\gamma}{d\Omega}}}}

\newcommand{\gn}{\ensuremath{\gamma N}}
\newcommand{\gp}{\ensuremath{\gamma p}}
\newcommand{\pn}{\ensuremath{\pi N}}
\newcommand{\en}{\ensuremath{\eta N}}
\newcommand{\pD}{\ensuremath{\pi \Delta}}
\newcommand{\sn}{\ensuremath{\sigma N}}
\newcommand{\rn}{\ensuremath{\rho N}}
\newcommand{\on}{\ensuremath{\omega N}}
\newcommand{\ppn}{\ensuremath{\pi\pi N}}

\newcommand{\kl}{\ensuremath{K\Lambda}}
\newcommand{\ks}{\ensuremath{K\Sigma}}

\begin{document}

\preprint{JLAB-THY-08-801}

\title{Dynamical coupled channel calculation of pion and omega meson production
\footnotetext{Notice: Authored by Jefferson Science Associates, LLC under 
U.S. DOE Contract No. DE-AC05-06OR23177. The U.S. Government retains a 
non-exclusive, paid-up, irrevocable, world-wide license to publish or 
reproduce this manuscript for U.S. Government purposes.}}

\author{Mark W.\ Paris}
\email[]{mparis@jlab.org}
\affiliation{Excited Baryon Analysis Center,
Thomas Jefferson National Accelerator Facility,
12000 Jefferson Avenue MS12H2,
Newport News, Virginia, 23606}


\date{\today}

\begin{abstract}
A dynamical coupled channel approach is used to study
$\pi$ and $\omega$--meson production induced by pions and photons scattering 
from the proton. Six intermediate channels including 
\pn, \en, \pD, \sn, \rn, \on\ are employed to describe
unpolarized and polarized data. Bare parameters in an effective 
hadronic Lagrangian are determined in a fit to the data for
\pnpn, \gnpn, \pmpon, and \gpop\ reactions at center-of-mass 
energies from threshold to $W < 2.0$ GeV. The $T$ matrix
determined in these fits is used to calculate the photon beam 
asymmetry for $\omega$-meson production and the \onon\ total cross
section and \on\ scattering lengths.
\end{abstract}


\maketitle

\section{Introduction}
Nucleon resonances are thought to play a decisive role in reactions of
strong, electromagnetic and weak probes on nucleons at energies $W<2.0$ GeV.
The extent to which nucleon resonances determine unpolarized and
polarized observables in meson production reactions and the role of
non-resonant contributions in these reactions remains an open question 
in this energy region. Model determinations of the $T$
matrix consistent with the observed meson production data in this 
kinematic regime seeks to resolve the resonance spectrum of 
the nucleon. Such a determination offers the possibility of gaining
insight into fundamental aspects of quantum chromodynamics, such as 
the role of chiral symmetry,
confinement and a detailed understanding of the correlations among
the strongly interacting quarks.

Limited experimental data for $\omega$ meson production in the
resonance region is being rapidly augmented. There is existing high 
precision data from the {\sc SAPHIR} collaboration \cite{Barth:2003kv} 
for the \gpop\ reaction from which the unpolarized differential cross
section (DCS) and decay angular distributions have been extracted. Consistent 
with this data are the more recent measurements of the GRAAL 
collaboration \cite{Hourany:2005wh,Ajaka:2006bn}. More photoproduction 
data at similar kinematics is anticipated from the CLAS 
collaboration \cite{Williams:2005em}. The \pnon\ data from bubble and 
drift chamber experiments is of low precision and there is little 
overlap in different experiment's \cite{Danburg:1971ui,Karami:1979ib}
kinematics. Though there is some 
discussion in the literature about the validity of the extracted 
cross sections \cite{Hanhart:1998wx,Hanhart:2001ft} we assume the
data is correct as originally published.

The importance of including off-shell effects in dynamical coupled 
channel formulations of strong and electromagnetic meson production 
reactions
has been extensively studied \cite{Yang:1985yr,*Nozawa:1990vd,*Lee:1991dd,
Krehl:1999km}.
The present study incorporates off-shell effects in a coupled channel
approach and is comparable to the model treatments of Krehl {\em et.\ al.}
\cite{Krehl:1999km} and Chen {\em et.\ al.} \cite{Chen:2007cy}.
It should be contrasted with coupled channel calculations 
which take into account coupling of the intermediate states
only to the continuum and neglect their
off-shell contributions such as those of the Giessen
group \cite{Shklyar:2004ba} and KVI \cite{Usov:2005wy}. In the Giessen 
study, an effective Lagrangian is adopted for the channels \pn, \ppn, \en,
\on, \kl, \ks. They assume a resonant contribution similar to the
one adopted in the present study.

Motivations for studying the \on\ reaction are manifold. Besides the
insight into the $T=\shalf$ resonance spectrum and implications for
meson production reactions, the vector mesons are thought to be 
important components in very dense matter in the neutron-rich stellar
environment\cite{Brown:1991kk,Trnka:2005ey,Muehlich:2006nn}. 
In nuclear matter, the $\omega\nn$ coupling can play a 
large role in determining the equation of state in some models
\cite{Tuchitani:2004ur}.

Any realistic model reaction theory for meson production necessarily
incorporates a large number of parameters. In the 
non-resonant terms, we require intermediate state particle masses, 
bare couplings and cutoffs of the hadronic degrees of freedom.
For the resonant terms we need the bare masses, bare couplings, and 
cutoffs for each resonance in a given partial wave. Given this
large number of parameters required to fully determine the
$T$ matrix the question arises as to the physical and predictive content
of such a model. In order to address this question in the present study 
we take the following approach. We determine the $T$ matrix by fitting
a subset of the available date (described in detail in 
Section \ref{sec:results}) and using this, calculate another, unfitted 
observable (here the photon beam asymmetry for $\omega$ meson production
shown in Fig.\ref{fig:gosig}).
The quality of the prediction of the unfitted observable is a measure of
the utility of model in determining more detailed information such as
polarization observables from less detailed ones, like unpolarized total
or differential cross sections. Deficiencies of such a predicted fit
indicate the need to improve the model assumptions and 
to augment the included dynamics.

In the next section we briefly describe the model theory for the 
six-channel model. The results of the fit to the data are presented 
and discussed in Section \ref{sec:results}. The final section gives
conclusions and descriptions for improvements to the present study which
are under development and outlines possible applications of the present
approach to other reactions including $\rho$ and $\phi$ production,
pion and $\omega$ meson electroproduction and reactions on nuclear targets.

\begin{figure}[t]
\includegraphics[ width=250 pt, keepaspectratio, clip]{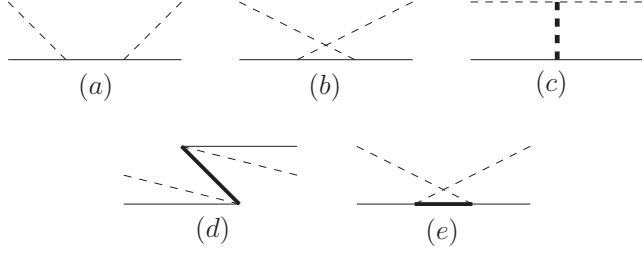}
\caption{\label{fig:vpnpn}The interactions \vmbmb\ and \vmbgn\ include
74 interaction mechanisms. In this figure and Figs.\ref{fig:vonpn},
\ref{fig:vonon} and \ref{fig:vgn} we show a subset of these. 
Here the \vpnpn\ interaction mechanisms are shown.
(a) $s$-channel nucleon exchange; (b) $u$-channel nucleon exchange; (c) 
$t$-channel $\rho$ exchange;
(d) $s$-channel $\bar{\Delta}$ exchange; (e) $u$-channel $\Delta$ exchange.}
\end{figure}
\begin{figure}[t]
\includegraphics[ width=250 pt, keepaspectratio, clip]{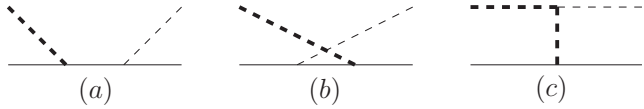}
\caption{\label{fig:vonpn}The \vonpn\ interaction mechanisms.
(a) $s$-channel nucleon exchange; (b) $u$-channel nucleon exchange; 
(c) $t$-channel $\rho$ exchange.}
\end{figure}
\begin{figure}[t]
\includegraphics[ width=170 pt, keepaspectratio, clip]{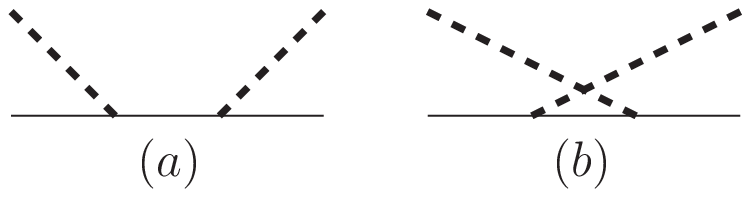}
\caption{\label{fig:vonon}The \vonon\ interaction mechanisms.
(a) $s$-channel nucleon exchange; (b) $u$-channel nucleon exchange.}
\end{figure}

\section{Model reaction theory\label{sec:model}}
The $T$ matrix for \pnmb\ and \gnmb\ -- in this work the final $MB$
state is restricted to \pn,\on\ -- is written as a sum of non-resonant, 
$t$ and resonant, $t^R$ contributions
\begin{align}
T(E) = t(E) + t^R(E),
\end{align}
where $E=W$ is the scattering energy of the particles in the center-of-mass
frame. Qualitatively, the non-resonant contribution 
includes rescattering and
coupled channel effects on the Born amplitudes while the resonant
contribution includes these effects on the bare resonance transition
form factors and bare resonance masses.
No assumption is made about the relative size of the contributions of
these terms. Our first objective in this work is to determine the $T(E)$ in
fits to the observed data.

\begin{table}[htbp]
\begin{tabular*}{0.75\textwidth}{@{\extracolsep{\fill}}c|c|c|cc|c|ccc|ccc}
$\ell_{TJ}$ & \multicolumn{1}{c}{$\pi N$} & \multicolumn{1}{c}{$\eta N$} & \multicolumn{2}{c}{$\pi \Delta$} & \multicolumn{1}{c}{$\sigma N$} & \multicolumn{3}{c}{$\rho N$} & \multicolumn{3}{c}{$\omega N$} \\
\hline
$S_{11}  $&$(0,\shalf) $&$(0,\shalf) $&$(2,\sthalf)$&$           $&$(1,\shalf) $&$(0,\shalf) $&$(2,\sthalf)$&$           $&(0,\shalf) & (2,\sthalf) & $  $        \\
$S_{31}  $&$(0,\shalf) $&$           $&$(2,\sthalf)$&$           $&$           $&$(0,\shalf) $&$(2,\sthalf)$&$           $&$           $&$           $&$           $ \\
$P_{11}  $&$(1,\shalf) $&$(1,\shalf) $&$(1,\sthalf)$&$           $&$(0,\shalf) $&$(1,\shalf) $&$(1,\sthalf)$&$           $&(1,\shalf) & (1,\sthalf) & $  $        \\
$P_{13}  $&$(1,\shalf) $&$(1,\shalf) $&$(1,\sthalf)$&$(3,\sthalf)$&$(2,\shalf) $&$(1,\shalf) $&$(1,\sthalf)$&$(3,\sthalf)$&(1,\shalf) & (1,\sthalf) & (3,\sthalf)  \\
$P_{31}  $&$(1,\shalf) $&$           $&$(1,\sthalf)$&$           $&$           $&$(1,\shalf) $&$(1,\sthalf)$&$           $&$           $&$           $&$           $ \\
$P_{33}  $&$(1,\shalf) $&$           $&$(1,\sthalf)$&$(3,\sthalf)$&$           $&$(1,\shalf) $&$(1,\sthalf)$&$(3,\sthalf)$&$           $&$           $&$           $ \\
$D_{13}  $&$(2,\shalf) $&$(2,\shalf) $&$(0,\sthalf)$&$(2,\sthalf)$&$(1,\shalf) $&$(2,\shalf) $&$(0,\sthalf)$&$(4,\sthalf)$&(2,\shalf) & (0,\sthalf) & (2,\thalf)   \\
$D_{15}  $&$(2,\shalf) $&$(2,\shalf) $&$(2,\sthalf)$&$(4,\sthalf)$&$(3,\shalf) $&$(2,\shalf) $&$(2,\sthalf)$&$(4,\sthalf)$&(2,\shalf) & (2,\sthalf) & (4,\sthalf)  \\
$D_{33}  $&$(2,\shalf) $&$           $&$(0,\sthalf)$&$(2,\sthalf)$&$           $&$(2,\shalf) $&$(0,\sthalf)$&$(2,\sthalf)$&$           $&$           $&$           $ \\
$D_{35}  $&$(2,\shalf) $&$           $&$(2,\sthalf)$&$(4,\sthalf)$&$           $&$(2,\shalf) $&$(2,\sthalf)$&$(4,\sthalf)$&$           $&$           $&$           $ \\
$F_{15}  $&$(3,\shalf) $&$(3,\shalf) $&$(1,\sthalf)$&$(3,\sthalf)$&$(2,\shalf) $&$(3,\shalf) $&$(1,\sthalf)$&$(3,\sthalf)$&(3,\shalf) & (1,\sthalf) & (3,\sthalf)  \\
$F_{17}  $&$(3,\shalf) $&$(3,\shalf) $&$(3,\sthalf)$&$(5,\sthalf)$&$(4,\shalf) $&$(3,\shalf) $&$(3,\sthalf)$&$(5,\shalf) $&(3,\shalf) & (3,\sthalf) & (5,\sthalf)  \\
$F_{35}  $&$(3,\shalf) $&$           $&$(1,\sthalf)$&$(3,\sthalf)$&$           $&$(3,\shalf) $&$(1,\sthalf)$&$(3,\sthalf)$&$           $&$           $&$           $ \\
$F_{37}  $&$(3,\shalf) $&$           $&$(3,\sthalf)$&$(5,\sthalf)$&$           $&$(3,\shalf) $&$(3,\sthalf)$&$(5,\sthalf)$&$           $&$           $&$           $ \\
\end{tabular*}
\caption{\label{tab:LS}The $(L,S)$ terms for partial waves $\ell_{TJ}$
for included channels.}
\end{table}

Except for the \pnpn\ reaction where we fit to energy dependent solution
of Arndt {\em et.\ al.} \cite{Arndt:2006bf} for the 
\pnpn\ partial wave amplitudes,
we fit the unpolarized and polarized cross 
sections of the $\gamma$ and $\pi$ induced reactions.
The DCS for $\pi$ induced reactions
is related to the $T$ matrix as
\begin{align}
\dspdo &= \frac{(4\pi)^2}{k^2} \rho_{M'B'}(k')\rho_{\pn}(k)
\half \sum_{M_{M'},M_{N'}}\sum_{M_N}
\left|T_{M_{M'}M_{N'},M_N}(\bvec{k}',\bvec{k};E)\right|^2,
\end{align}
where $\bvec{k}$ is the relative momentum of the initial \pn\ state
and $\bvec{k}'$ is the relative momentum of the final meson-nucleon ($M'N'$)
state, where $M'=\pi$ or $\omega$.
The spin projection of the particles in the inital (final) 
state is $M_N$ ($M_{M'},M_{N'}$). The quantity
$\rho_{MB}(p) = \frac{\pi p E_M(p) E_B(p)}{E_M(p)+E_B(p)}$, where
$E_i(p) = \sqrt{p^2+m_i^2}$ is related to the density of states.
A similar relation holds for the unpolarized photoproduction cross
section \dsgdo.

\begin{figure}[htbp]
\includegraphics[ width=230 pt, keepaspectratio, clip]{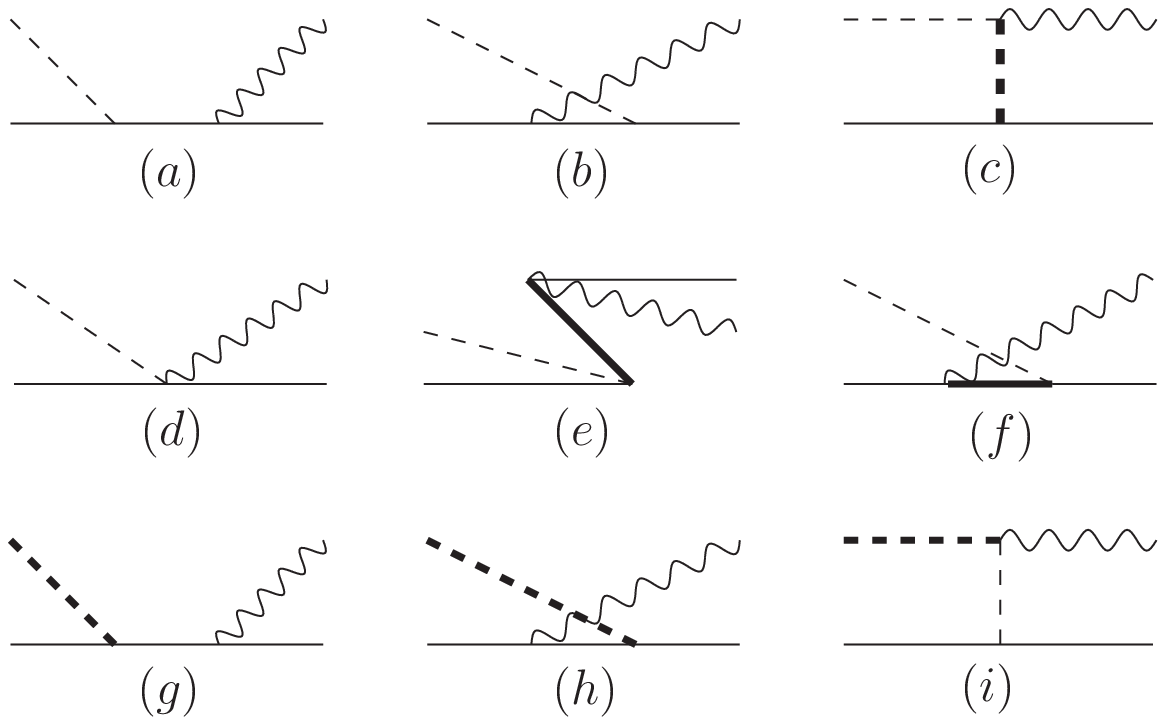}
\caption{\label{fig:vgn}The \vpngn\ and \vongn\ interaction
mechanisms. For photoproduction of $\pi$ mesons:
(a) $s$-channel nucleon; (b) $u$-channel nucleon; (c) $t$-channel $\pi$,
$\rho$, and $\omega$; (d) contact; (e) $s$-channel $\bar{\Delta}$;
(f) $u$-channel $\Delta$. For photoproduction of $\omega$ mesons:
(g) $s$-channel; (h) $u$-channel; (i) $t$-channel $\pi$ exchange.}
\end{figure}

\subsection{Non-resonant contribution\label{subsec:non-res}}
The non-resonant contribution to the transition matrix in the 
partial-wave representation for the pion-induced $\tjlsmnpn(E)$
and the mixed partial-wave/helicity representation for the
photon-induced $\tjlsmngn(E)$ reactions are
\begin{align}
\label{eqn:nr_strong}
&\tjlsmnpn(k',k;E) = \vjlsmnpn(k',k) \nonumber \\
&+ \sum_{LSMB} \int_0^\infty dp\,p^2 \tjlsmnmb(k',p;E) \Gmb(p;E) \vjlsmbpn(p,k),
\\
\label{eqn:nr_em}
&\tjlsmngn(k,q;E) = \vjlsmngn(k,q) \nonumber \\
&+ \sum_{LSMB} \int_0^\infty dp\,p^2 \tjlsmnmb(k,p;E) \Gmb(p;E) \vjlsmbgn(p,q),
\end{align}
where $T$ is the total isospin and $T_{N,z}$ is the isospin projection
of the nucleon in the initial state, 
$J$ is the total angular momentum of the partial wave, $L(L')$
is the partial wave orbital angular momentum of the initial(final) state,
$S(S')$ is the total intrinsic spin of the particles in the initial(final)
state and
$\ell = J\pm\shalf$ is the \pn\ initial state orbital angular momentum.
The included partial waves are shown in Table \ref{tab:LS}.
Channels are defined by
the meson species $M'(M)$ of the final (intermediate) state and $N(N')$,
the nucleon of the initial(final) state or $B$ the intermediate state 
baryon associated with the meson $M$ of channel $MB$. The sums, $\sum_{MB}$
are over channels \pn, \en, \pD, \sn, \rn, and \on.
Here $\Gmb(p;E)$ is the relativistic free particle Green's function 
\begin{align}
\label{eqn:GF}
\Gmb(p;E) &= \frac{1}{E-E_M(p)-E_B(p)-\Sigma_{MB}(p;E)}.
\end{align}
$E_i(p)$ are the free particle energies with masses given in 
Table \ref{tab:mass} and $\Sigma_{MB}(p;E)$ is the self-energy of the
unstable particle in channels $MB=\pD,\sn,\rn$ including the effects
induced by the decay of the unstable particle in these channels
\cite{MSL}. Channels with stable particles $MB=\pn,\en,\on$  have
$\Sigma_{MB} = -i\eps$ corresponding to the coupling to the on-shell
intermediate states.
The width of the $\omega$ meson
$\Gamma_\omega=8.5(1)$ MeV is neglected.

\begin{figure}[t]
\includegraphics[ width=350 pt, keepaspectratio, clip]{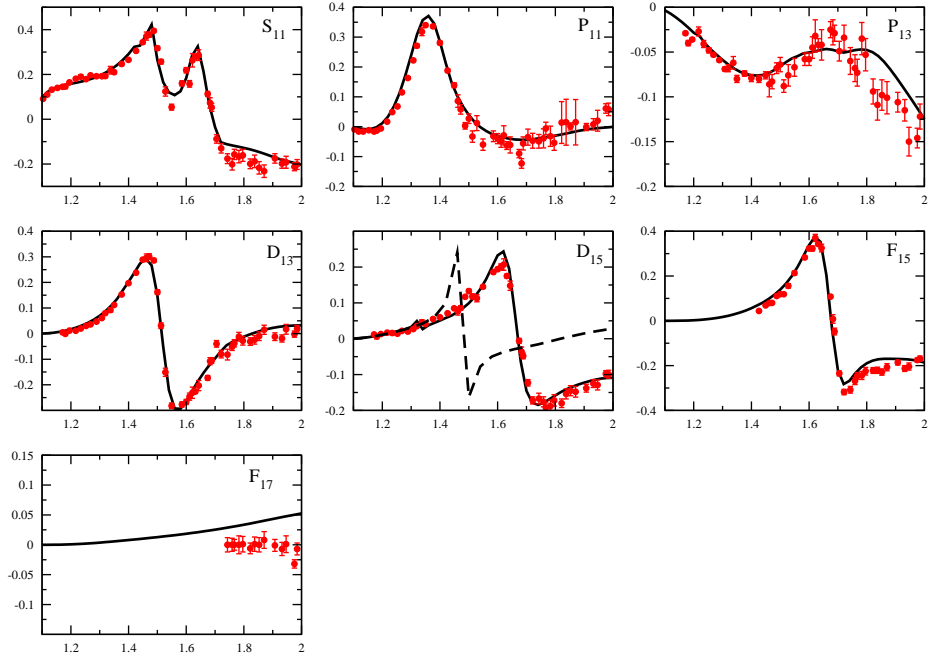}
\caption{\label{fig:rtpnpn1} Real part of \pnpn\ partial wave amplitudes 
$\overline{T}^{J1}_{\ell\pn,\ell\pn}(k_{\pn},k_{\pn};E)$
for $T=1/2$ versus center-of-mass energy, $W$ (in GeV) fit to
single energy extraction of Ref.\cite{Arndt:2006bf}. The
dashed line shows the best fit obtained without the second resonance
in $D_{15}$.}
\end{figure}
\begin{figure}[t]
\includegraphics[ width=350 pt, keepaspectratio, clip]{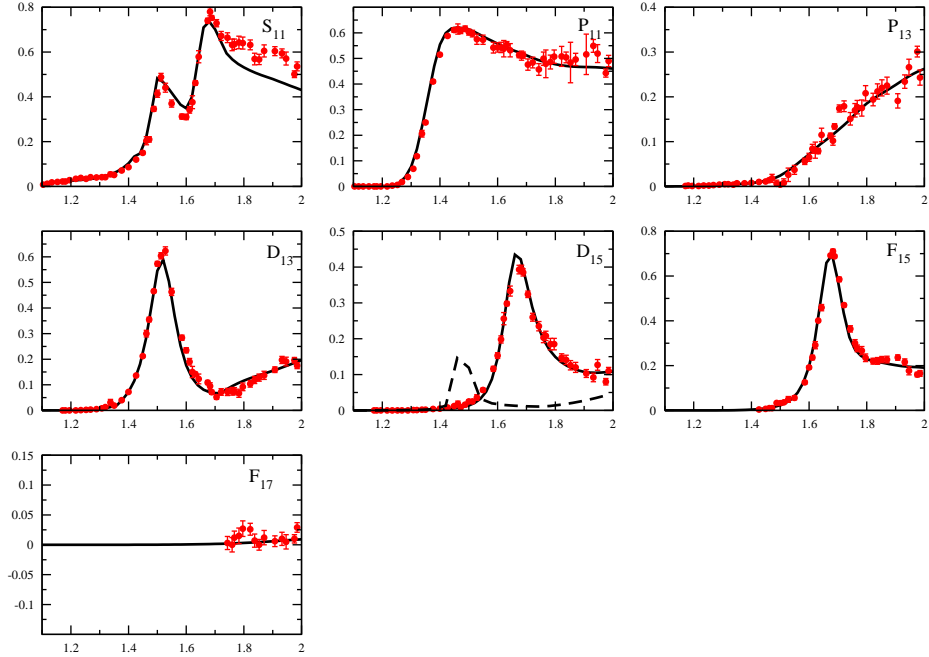}
\caption{\label{fig:itpnpn1} Imaginary part of \pnpn\ partial wave
amplitudes $\overline{T}^{J1}_{\ell\pn,\ell\pn}(k_{\pn},k_{\pn};E)$
for $T=1/2$ versus center-of-mass energy, $W$ (in GeV)
fit to single energy extraction of Ref.\cite{Arndt:2006bf}.
The dashed line shows the best fit obtained without the second resonance
in $D_{15}$.}
\end{figure}

The $\vmbmb$ and $\vmbgn$ are the effective non-resonant 
interaction Hamiltonians for hadrons 
$\pi$, $\eta$, $\sigma$, $\rho$, $\omega$, $N$, $\Delta$, and photons,
$\gamma$. These interactions are Born amplitudes derived 
from the Lagrangian of Ref.\cite{MSL} and
subjected to the unitary transformation method of Ref.\cite{SKO}.
It yields an interaction which is independent of the scattering energy,
$E=W$, and depends only on the relative three-momenta of the incoming 
and outgoing particles. 

In the present model there are 74 interaction mechanisms
among the channels \pn, \en, \pD, \sn, \rn, \on, and \gn.
Figures \ref{fig:vpnpn}--\ref{fig:vgn} show examples of
these explicitly for the terms involving \gn, \pn, and \on\ channels
for pion induced and photoproduction amplitudes. In this work we neglect 
the contribution of the high-mass $a_J$ and $f_J$ mesons (except
the $f_0/\sigma(600)$). We make the further simplification in the
non-resonant hadronic interaction involving the \on\ channel of including 
only those terms which couple the \on\ to itself and to \pn, that is:
$v_{\on,MB}=v_{\on,\pn} \delta_{MB,\pn}+v_{\on,\on} \delta_{MB,\on}$.
This simplification permits the introduction of a minimal number of 
additional bare parameters for the \on\ channel while retaining effects 
from each of the non-resonant $s-$, $t-$, and $u-$ mechanisms. In this 
way, we capture the behavior associated with each mechanism, while 
maintaining a tractable model. 

\begin{figure}[t]
\includegraphics[ width=350 pt, keepaspectratio, clip]{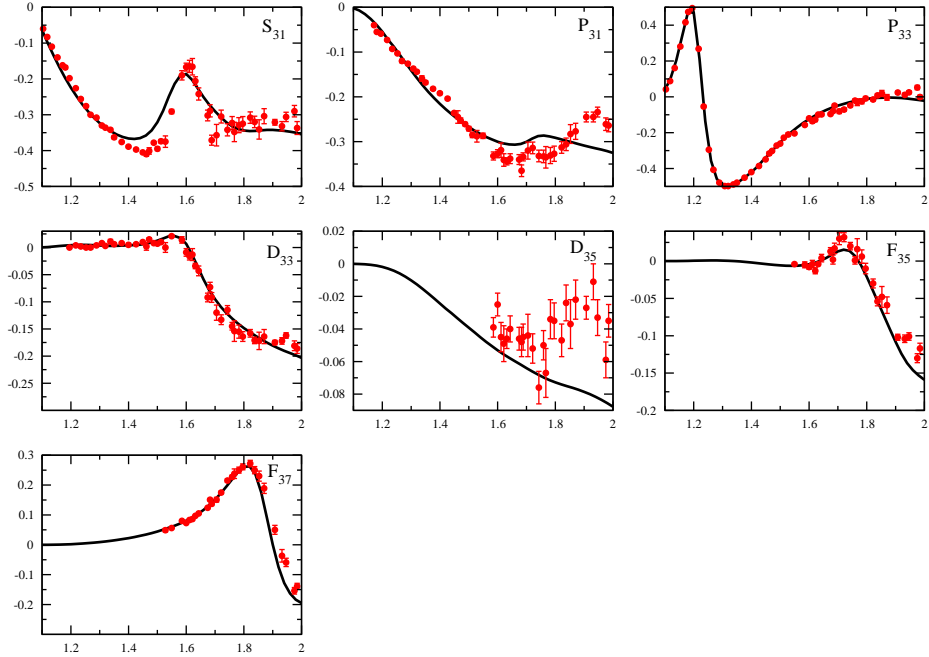}
\caption{\label{fig:rtpnpn3} Real part of \pnpn\ partial wave amplitudes 
$\overline{T}^{J3}_{\ell\pn,\ell\pn}(k_{\pn},k_{\pn};E)$
for $T=3/2$ versus center-of-mass energy, $W$ (in GeV) fit to
single energy extraction of Ref.\cite{Arndt:2006bf}.}
\end{figure}
\begin{figure}[t]
\includegraphics[ width=350 pt, keepaspectratio, clip]{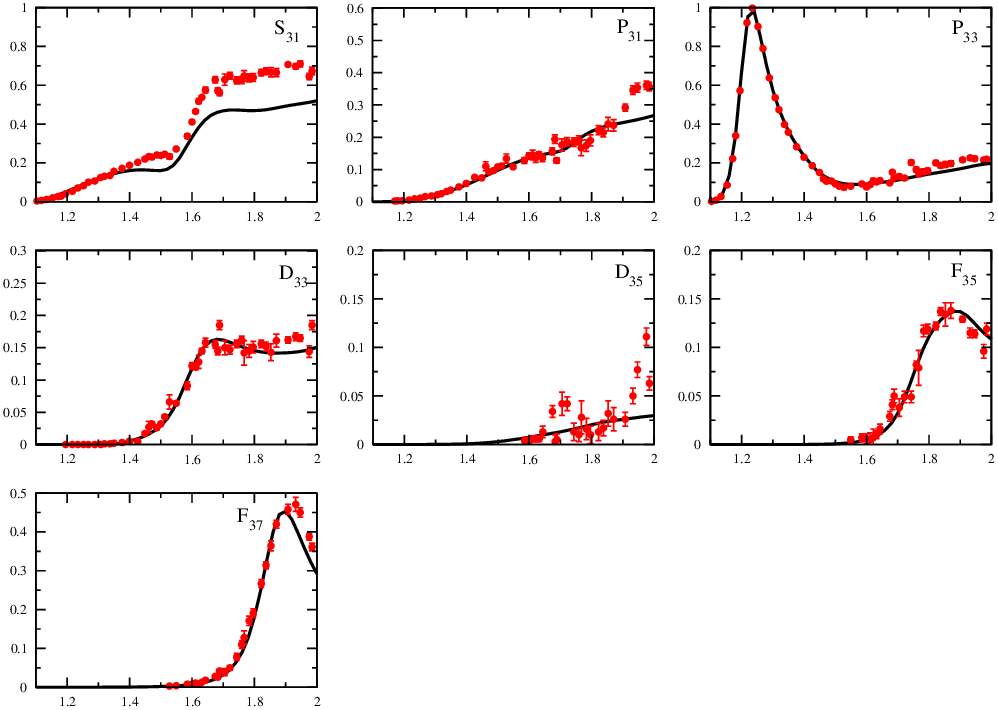}
\caption{\label{fig:itpnpn3} Imaginary part of \pnpn\ partial wave
amplitudes $\overline{T}^{J3}_{\ell\pn,\ell\pn}(k_{\pn},k_{\pn};E)$
for $T=3/2$ versus center-of-mass energy, $W$ (in GeV)
fit to single energy extraction of Ref.\cite{Arndt:2006bf}.}
\end{figure}

Equations \eqref{eqn:nr_strong} and \eqref{eqn:nr_em} represent the bulk
of the computational effort required to carry out the coupled channel
dynamical approach (at the two-body level). Most of the computer time 
required ($\sim 3/5$)
is spent evaluating the Born terms. Much of the remainder is spent
inverting the matrix representing the scattering wave function, 
$\mathcal{F}^{-1}=(1-vG_0)^{-1}=1+tG_0$ appearing in 
Eq.\eqref{eqn:nr_strong} on a momentum grid of 25 {\em Gauss-Legendre}
points using standard subtraction methods \cite{Tabakin:1970sa}
. Convergence has been checked with grids of up to 45 {\em GL} points.
A parallel {\sc fortran90} code has been developed to cope with the long
evaluation times for a single $\chi^2$ evaluation ($\sim 100$ node$\cdot$m).
It exploits the independence of the partial waves and energies in the
evaluation of the $T$ matrix. Typically $\sim 10^3-10^4$ $\chi^2$ 
evaluations are required for optimization using the {\sc minuit} 
package\cite{James:1975dr}.

The amplitudes $\tjlsmbpn$
include partial wave contributions up to and including
$L=3$ ($F$--wave). The same is true for the electromagnetic terms
except for the $t$-channel pion exchange in Fig.\ref{fig:vgn}(i). 
In this case, all partial waves are required for convergence. 
For $L>3$ the contribution to the electromagnetic non-resonant
amplitude $\tjlsmngn(k,q;E)$ is calculated at the Born amplitude
level only and neglects the effects due to final state interactions and 
coupled channels, {\em ie.}\ the second terms of Eq.\eqref{eqn:nr_em}.

\begin{figure}[htbp]
\includegraphics[ width=350 pt, keepaspectratio, clip]{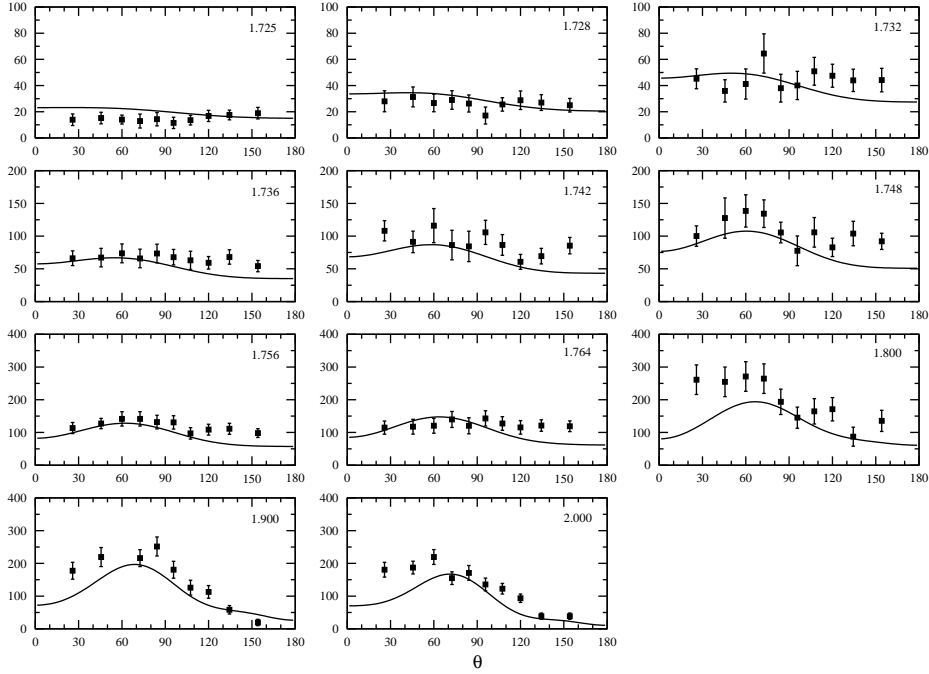}
\caption{\label{fig:podxs}Differential cross section vs.\ center-of-mass
angle, $\theta$ for \pmpon\ (in
mb/sr) compared to data from Refs.\cite{Karami:1979ib,Danburg:1971ui}. 
The center-of-mass energy, $W$ (in GeV) is shown in the upper-right
corner of each panel.}
\end{figure}

The non-resonant interaction depends on the masses of the hadrons and
their coupling and cutoff parameters. These values obtained in the
five-channel fit of Ref.\cite{MSL} are shown here in Tables \ref{tab:nr_c} 
and \ref{tab:nr_l} for completeness. For the interaction terms, \vmbmb\ 
and \vmbgn\ (other than the mass of $f_0/\sigma(600)$ which is a fit
parameter), the physical particle 
masses are used. Form factors are included at vertices in the non-resonant 
interactions, \vmbmb\ and \vmbgn\ have the form
$F(\qmag;m)=(\Lambda^2/(\Lambda^2+\qmag^2))^m$. Here $\qmag$ is either
the momentum transferred at the vertex or the relative 
momentum \cite{Sato:1996gk}. We use the value $m=2$ at all vertices.

\begin{figure}[t]
\includegraphics[ width=350 pt, keepaspectratio, clip]{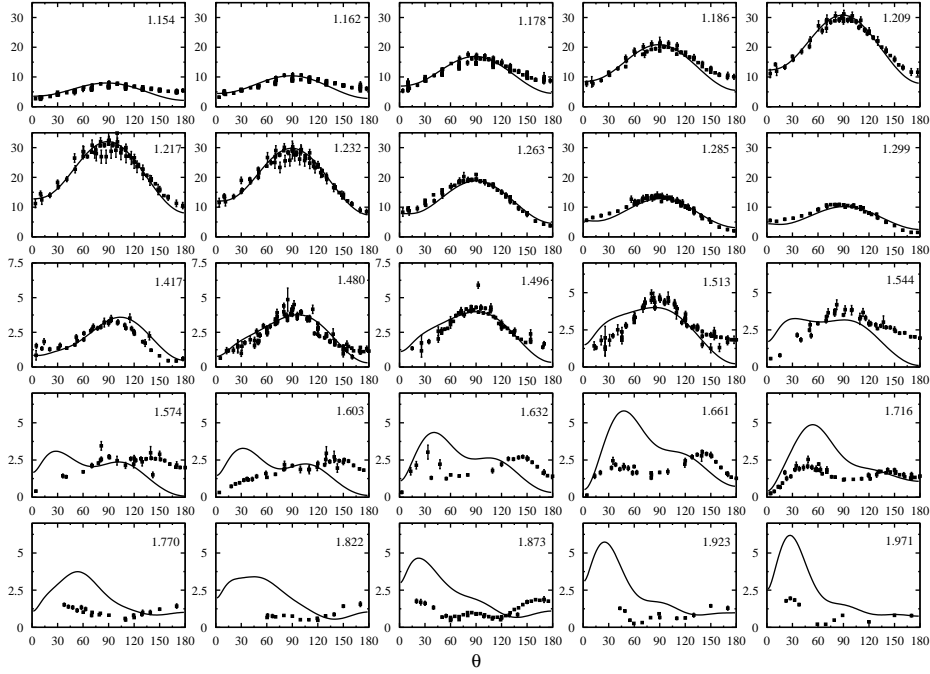}
\caption{\label{fig:g0dxs}
Unpolarized differential cross section vs.\ scattering angle $\theta$ 
in the center-of-mass system for \gppzp\ reaction compared to data
from Refs.
\cite{Fuchs:1996ja,*Genzel:1973jy,*Argan:1975pi,*Aleksandrov:1978ex,
*Dougan:1977zw,*Aleksandrov:1977tj,*Bartholomy:2004uz,*Dougan:1975pa,
*Yoshioka:1980vu,*Hemmi:1973nf,*Worlock:1960pi,*Feller:1974yd,*Hemmi:1973ii,
*Ahrens:2002gu,*Krusche:1999tv,*Althoff:1979mc,*Althoff:1979mb,
*Delcourt:1969dp,*DeStaebler:1965sf,*Dugger:2007bt,*Barton:1974sf,
*Booth:1974ms,*Abrahamian:1975vs,*Husmann:1977ar,
*Brefeld:1975dv,Abrahamian:1974jc,Arai:1977kb,Beck:1997ew, Ahrens:2004pf}. 
The center-of-mass energy, $W$ (in GeV) is shown in the upper-right 
corner of each panel.}
\end{figure}
\begin{figure}[t]
\includegraphics[ width=350 pt, keepaspectratio, clip]{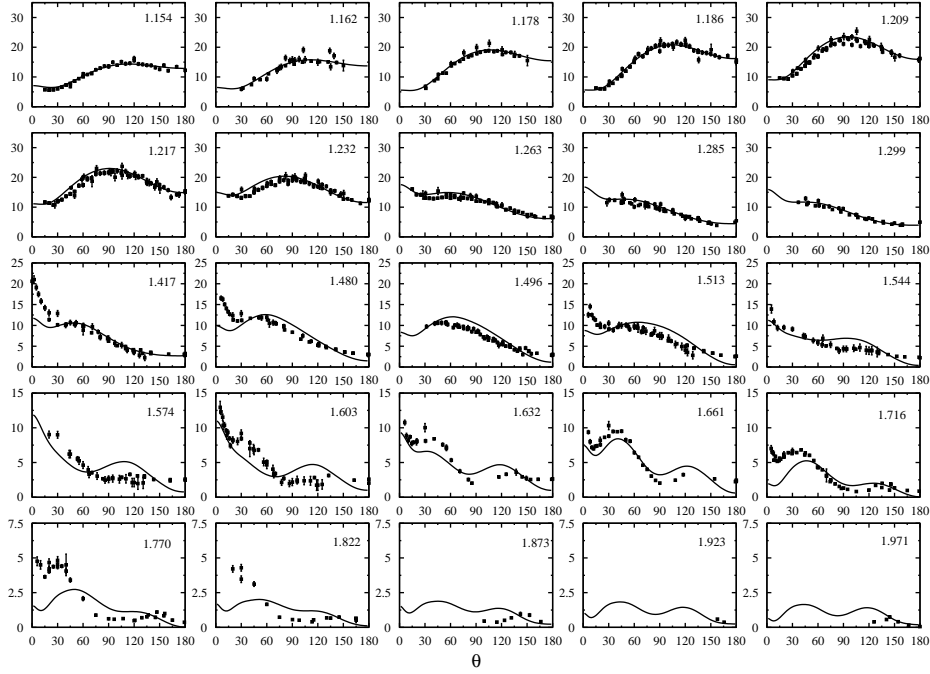}
\caption{\label{fig:gpdxs} 
Unpolarized differential cross section vs. $\theta$ 
for \gpppn\ reaction compared to data from Refs.
\cite{Fischer:1970df,*Fischer:1972mt,*Buechler:1994jg,
*Branford:1999cp,*Fujii:1976jg,*Fujii:1971qe,*Dannhausen:2001yz,
*Betourne:1968bd,*Aleksandrov:1970xq,*Faure:1984qe,*Althoff:1983te,
*PhysRev.159.1195,*Ahrens:2006gp, *Ekstrand:1972rt,
*PhysRevD.1.1946,*Zhu:2004dy, *PhysRevLett.17.1027,
Abrahamian:1974jc,Arai:1977kb,Ahrens:2004pf}. The center-of-mass
energy, $W$ (in GeV) is shown in the upper-right corner of each panel.}
\end{figure}
\begin{figure}[t]
\includegraphics[ width=350 pt, keepaspectratio, clip]{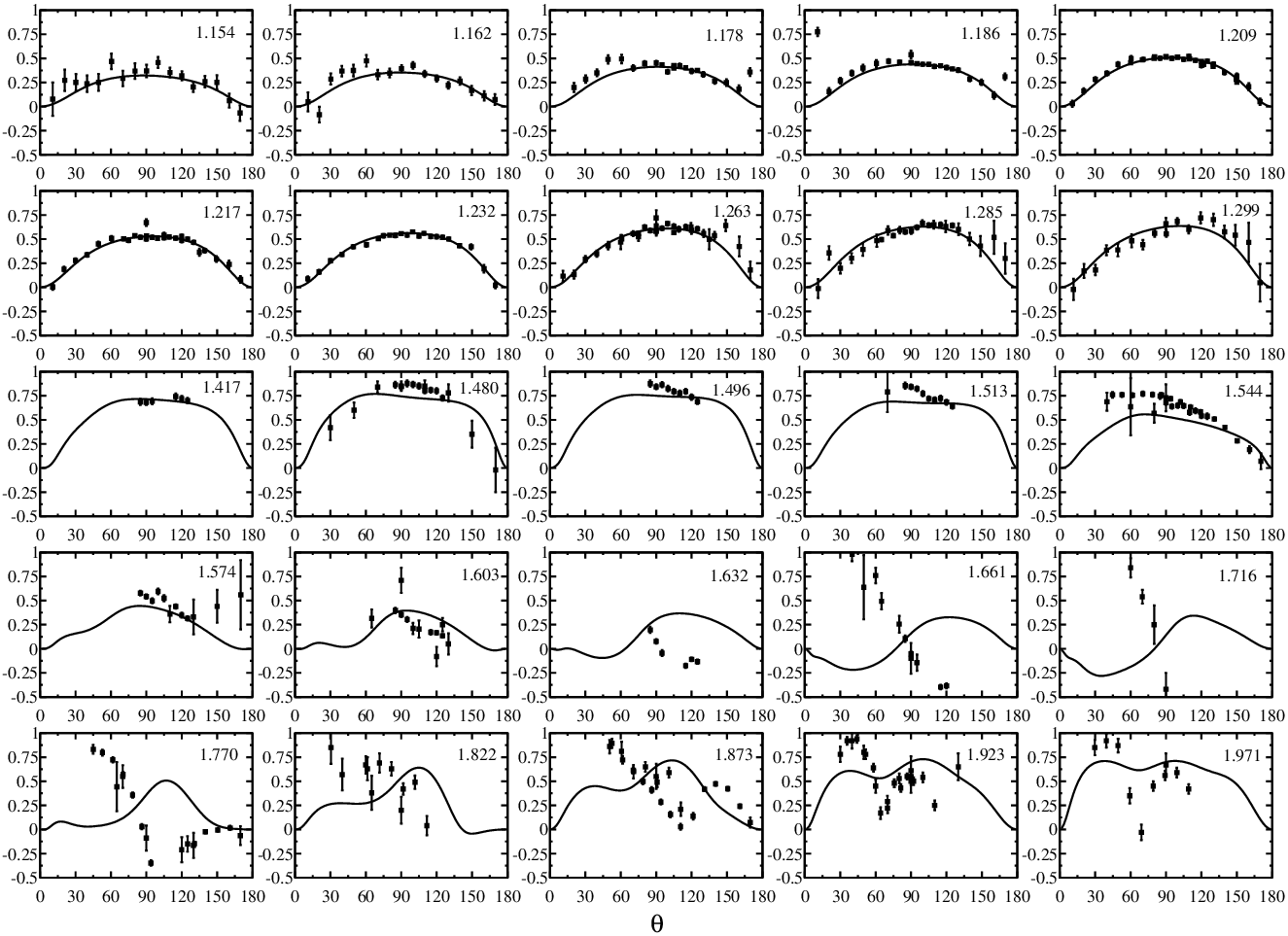}
\caption{\label{fig:g0sig} 
Photon beam asymmetry, $\Sigma_0(\theta,E)$ vs. $\theta$ for \gppzp\ 
reaction compared to data from Refs.
\cite{Blanpied:1992nn,*Belyaev:1983xf,*Barbiellini:1970qu,*Adamian:2000yi,
*Bussey:1975mt,
Knies:1974zx,Bussey:1979wt,Beck:1997ew,Blanpied:2001ae}. The center-of-mass
energy, $W$ (in GeV) is shown in the upper-right corner of each panel.}
\end{figure}
\begin{figure}[t]
\includegraphics[ width=350 pt, keepaspectratio, clip]{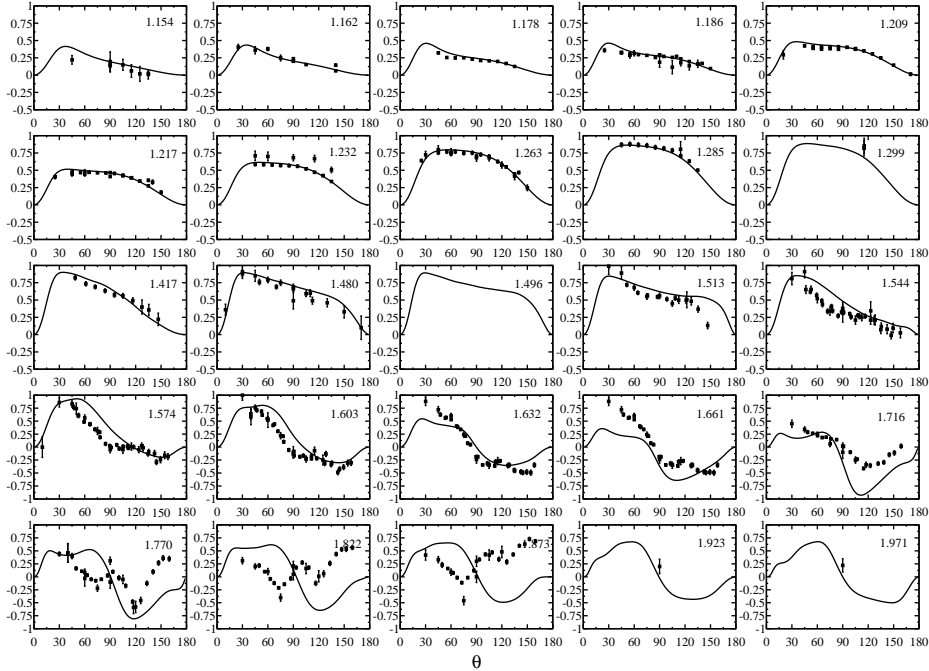}
\caption{\label{fig:gpsig} 
Photon beam asymmetry, $\Sigma_+(\theta,E)$ vs. $\theta$ for \gpppn\
reaction compared to data from Refs.
\cite{Zabaev:1975ag,*PhysRev.130.2429,*PhysRev.117.835,*Ganenko:1976ri,
*Getman:1981qt,*Ajaka:2000rj,*Alspector:1972pw,
Abrahamian:1974jc,Knies:1974zx,Bussey:1979wt,Beck:1997ew,Blanpied:2001ae}.
The center-of-mass energy, $W$ (in GeV) is shown in the upper-right corner
of each panel.}
\end{figure}
\begin{figure}[ht]
\includegraphics[ width=350 pt, keepaspectratio, clip]{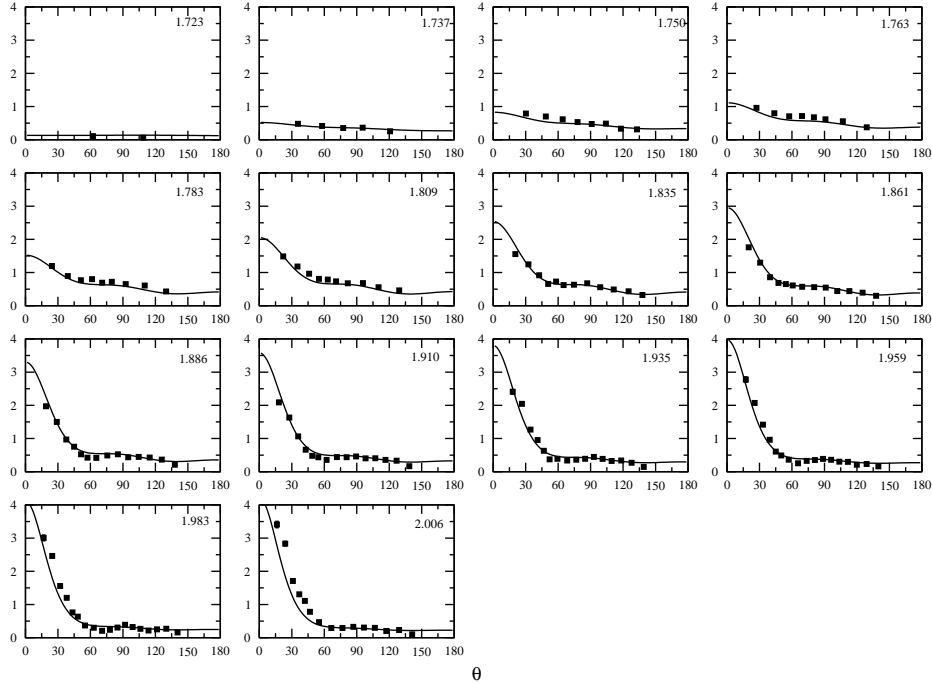}
\caption{\label{fig:godxs}Unpolarized differential cross section 
for \gpop\ versus $\theta$ (in $\mu$b/sr) compared to data from 
Ref.\cite{Barth:2003kv}. The center-or-mass energy, $W$ (in GeV) is
shown in upper-right of each panel.}
\end{figure}

\subsection{Resonant contribution\label{subsec:res}}
The resonant contribution, $t^R(E)$ to the scattering matrix is given as
\begin{align}
\Tjlsmbpn(k',k;E) &= \sum_{i,j}\Gjlsi(k';E) \Drij(E) \Gjpn(k;E), \\
\Tjlsmbgn(k,q;E)  &= \sum_{i,j}\Gjlsi(k ;E) \Drij(E) \Gjem(q;E),
\end{align}
where the sums $\Sigma_{i,j}$ run over the resonances in a given partial
wave (at most two per channel in this work) and $\Gbar$ is the dressed 
vertex function
\begin{align}
\Gjlsi(k;E) &= 
\Gbjlsi(k) \nonumber \\ &+ \sum_{L'S'M'B'} 
\int\! dk'\,k'^2 \tjlsmbmb(k,k';E) G_{0,M'B'}(k';E) \Gbjlspi(k') \\
\Gign(q;E) &=
\Gbign(q) \nonumber \\ &+ \sum_{LSMB} 
\int\! dk\,k^2 \Gijls(k;E) G_{0,MB}(k;E) \vjlsmbgn(k,q) \\
\end{align}
and $\Drij(E)$ is the dressed resonance propagator which depends on
the resonance self-energy, $\Sigma_{ij}(E)$:
\begin{align}
\mathcal{D}_{ij}(E)  &= (E-M^{(0)}_{N^*_i})\delta_{ij} - \Sigma_{ij}(E) \\
\Sigma_{ij}(E) &= \sum_{LSMB} \int\!dk\, k^2 
\Gbijls(k;E) G_{0,MB}(k;E) \Gjlsj(k;E).
\end{align}

The bare vertex functions $\Gamma$ should, in principle, be calculated 
from appropriate {\em ab initio} hadronic models. This is beyond the 
scope of the present study. Instead, we parameterize the vertex function 
in the center-of-mass for the partial wave specified by $J,L,S$ for
the hadronic channels as
\begin{align}
\label{eqn:Gmb}
\Gamma_{LSMB,N^*}^{JT}(k)
&=\zeta_{MB} \frac{1}{(2\pi)^{3/2}}\frac{1}{\sqrt{m_N}}
C^{JT}_{N^*LSMB} \left(\frac{k}{m_\pi}\right)^L f^{JT}_{N^*LSMB}(k).
\end{align}
Here $\zeta_{MB}=-i$ for $MB=\pn,\en,\pD$ and $\zeta_{MB}=1$ 
for $MB=\sn,\rn,\on$. At small values of the relative $MB$ momentum
$k$, $\Gamma_{LSMB,N^*}^{JT}(k)$ has the form appropriate to the
threshold production behavior, $k^L$. It is regulated at large $k$
by the form factor, $f^{JT}_{N^*LSMB}(k)$, described below.

The bare electromagnetic coupling for $N^*\to\gn$
for all $N^*$ except the first $P_{33}$ resonance (number `8' in Tables
\ref{tab:resc} and \ref{tab:resl}) is given by
\begin{align}
\label{eqn:Ggn}
\Gamma_{N^*,\lag\lN T_{N,z}}^{JT}(q)
&= \frac{1}{(2\pi)^{3/2}}\sqrt\frac{m_N}{E_N(q_R)}
\,A^{JT}_{\lambda T_{N,z}}\,
\sqrt\frac{q_R}{q} \,
g^{JT}_{N^*\lambda T_{N,z}}(q) \,
\delta_{\lambda,\lag-\lN},
\end{align}
with $\Gamma_{N^*,-\lag,-\lN T_{N,z}}^{JT}=(-1)^{J+\ell+1/2} 
\Gamma_{N^*,\lag\lN T_{N,z}}^{JT}$ where $N^*$ is in partial wave
$\ell_{JT}$.
The form for the first $P_{33}$ resonance is shown in the 
Appendix. The photon momentum at the resonance threshold, $q_R$ is
$M^{(p)}_{N^*}=q_R+E_N(q_R)$ where the resonance mass is taken from the
{\em Review of Particle Properties}\cite{PDG}. The isospin projection of
the initial nucleon is $T_{N,z}$ and the helicities are $\lag$ and $\lN$.
We assume the forms $f^{JT}_{N^*LSMB}(k)=
\left[{{\Ljtlsmbn}^2}/({{\Ljtlsmbn}^2+(k-k_{N^*})^2})\right]^{L+2}$ and
$g^{JT}_{N^*\lambda T_{N,z}}(q)=1$.
The $C^{JT}_{N^*LSMB}$, $\Ljtlsmbn$, $k_{N^*}$ and A$^{JT}_{\lambda T_{N,z}}$
are fit parameters. 

\section{Results and discussion\label{sec:results}}
The first objective of the present study is the simultaneous description
of the pion and photon induced single pion and omega meson production 
data in a coupled channel approach. 
Recent high precision measurements of $\omega$ photoproduction make it
possible to strongly constrain coupled channel model reaction theories.
The DCS and $\omega$ spin density matrix elements (SDME) have been measured
at SAPHIR and published \cite{Barth:2003kv} and measured by the CLAS 
collaboration\cite{Williams:2005em,Williams:2007phd}.
The only other observable measured is the single polarization observable,
the photon beam asymmetry (PBA) $\Sigma_\omega(\theta,E)$ at 
GRAAL\cite{Hourany:2005wh,Ajaka:2006bn}. We have elected
to also include the older $\pi$ induced reaction data from threshold
($\sim 1.72$ GeV) to 1.764 GeV from the Nimrod 
synchrotron \cite{Karami:1979ib} and the Alvarez detector data from
1.75 GeV to 2.05 GeV \cite{Danburg:1971ui} (in 100 MeV bins).

The world data for pion photoproduction measurements of DCS, shown in
Figs.\ref{fig:g0dxs},\ref{fig:gpdxs} and PBA shown in
Figs.\ref{fig:g0sig},\ref{fig:gpsig} are obtained from the 
George Washington
University Center for Nuclear Studies Data Analysis Center\cite{SAID}.
These high precision data in \gppzp\ were taken from Refs.
\cite{Fuchs:1996ja,Abrahamian:1974jc,Arai:1977kb,Beck:1997ew,Ahrens:2004pf}
for DCS and Refs.\cite{Blanpied:1992nn,Knies:1974zx,Bussey:1979wt,
Beck:1997ew,Blanpied:2001ae} for PBA and in \gpppn\ from 
Refs.\cite{Fischer:1970df,Abrahamian:1974jc,Arai:1977kb,Ahrens:2004pf}
(\cite{Zabaev:1975ag,Abrahamian:1974jc,Knies:1974zx,Bussey:1979wt,
Beck:1997ew,Blanpied:2001ae}) for DCS(PBA), respectively.
For the purpose of $\chi^2$ optimization of the data with respect to
the bare parameters of the theory a truncated data set of the highest
precision data was used which covers as much of the angular range as
was available on a per energy bin basis.

\begin{figure}[ht]
\includegraphics[ width=300 pt, keepaspectratio, clip]{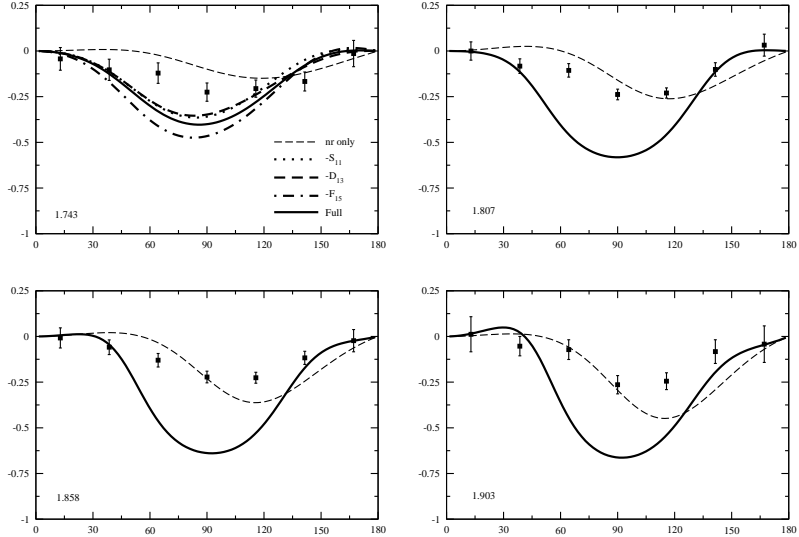}
\caption{\label{fig:gosig}Predicted photon beam asymmetry, 
$\Sigma_\omega(\theta,E)$ for \gpop\ (solid curve) for $E=W$
shown in lower-left corner of each panel compared with data from
Ref.\cite{Hourany:2005wh}. At the lowest energy,
the effect of removal of various resonances is shown. Removing
all (`nr only' -- thin dashed) and $S_{11}$ (dotted), $D_{13}$ (dashed)
, $F_{15}$ (dot-dashed).}
\end{figure}

In order to accomplish our objective we take as the starting point for
this analysis the $T$ matrix determined in
Ref.\cite{JuliaDiaz:2007kz} which fits the \pnpn\ partial wave amplitudes
\begin{align}
\overline{T}^{JT}_{\ell\pn,\ell\pn}(k_{\pn},k_{\pn};E)
&= -\rho_{\pn}(k_{\pn}) T^{JT}_{\ell\pn,\ell\pn}(k_{\pn},k_{\pn};E),
\end{align}
extracted from observed data by Ref.\cite{Arndt:2006bf} 
in the region 1.1 GeV $<E<$ 2.0 GeV in a {\em five}-channel approach, 
excluding \on. The non-resonant parameters are fixed and shown
in Tables \ref{tab:mass}, \ref{tab:nr_c}, and \ref{tab:nr_l}. 
The resonance parameters (\Mbres, \Cjtnlsmb, \Ljtnlsmb and \knstar) for
coupling to hadronic channels $MB=\{\pn,\en,\pD,\sn,\rn\}$ 
shown in Tables \ref{tab:resc} and \ref{tab:resl}.

Determination of the six-channel $T$ matrix \Tjlsmbmb, \Tfjlsmbgn
proceeds in two stages. At the first stage, the \pnpn\ $T=1/2$
partial wave amplitudes of Figs.\ref{fig:rtpnpn1} and \ref{fig:itpnpn1},
the \pmpon\ DCS of Fig.\ref{fig:podxs} and the \gpop\ DCS of 
Fig.\ref{fig:godxs} are fit simultaneously. This is accomplished by 
adjusting the non-resonant couplings $g^t_{\omega NN}$, 
$\kappa^t_{\omega NN}$ and $\Lambda^t_{\omega NN}$ appearing in the
$s-$ and $u-$channel $\omega$ emission and absorption of 
Figs.\ref{fig:vonpn}(a),(b), Figs.\ref{fig:vonon} and 
Figs.\ref{fig:vgn}(g),(h) and by adjusting the resonance 
parameters $N^*\to\on$, $G^{J1}_{N^*LS\on}$ and
$\Lambda^{J1}_{N^*LS\on}$. The introduction of the \on\ channel to the
calculation requires the addition of a second $D_{15}$ resonance,
shown in bold type in Tables \ref{tab:resc} and \ref{tab:resl}, in order
to fit the data. These points will be discussed in more detail below.

At the second stage of the fit, all non-resonant and hadronic channel
resonant parameters are fixed and the single meson photoproduction data
is fitted. Pion photoproduction data used for the fit includes the 
DCS in Figs.\ref{fig:g0dxs} and \ref{fig:gpdxs} and the PBA 
in Figs.\ref{fig:g0sig} and \ref{fig:gpsig} in the region
1.1 GeV $<E<$ 2.0 GeV. Omega meson photoproduction data used for
the fit includes only the SAPHIR measurement \cite{Barth:2003kv}
of the DCS from threshold, 1.72 GeV to 2.0 GeV shown 
in Fig.\ref{fig:gosig}. This is accomplished by
varying the photon helicity couplings, $A^{JT}_{N^*\lambda T_{N,z}}$
for $\lambda=\shalf,\sthalf$ and $T_{N,z}=+\shalf$. The resulting fits 
compared to the existing world data are shown as solid curves in
Figs.\ref{fig:g0dxs}--\ref{fig:gpsig}.


The overall quality of the fits to the complete set of data are in
fair agreement for energies $E<1.65$ GeV. The $T=1/2$ \pnpn\ partial 
wave amplitudes of Fig.\ref{fig:rtpnpn1} agree at the $1-\sigma$ 
level for all partial waves except the two highest. The $T=3/2$ partial
fits are of similar quality except for the $S_{11}$ wave and the $P_{11}$
wave at energies $E\gtrsim 1.9$ GeV. The fits to the photoproduction data
are good at low energies but degrade significantly at $E>1.65$ GeV
especially in the \gppzp\ reaction. Coupling to the $\pi\pi\! N$ channel
is expected to be large here.

\begin{table}[htbp]

\begin{minipage}{2in}
\begin{tabular}{cr}
\hline
$m_N             $&  938.5 \\
$m_\pi           $&  138.5 \\
$m_\eta          $&  547.5 \\
$m_\Delta        $& 1300.0 \\
$m_\sigma        $&  898.6 \\
$m_\rho          $&  811.7 \\
$m_\omega        $&  782.6 \\
\hline
\end{tabular}
\caption{\label{tab:mass}Propagator masses (MeV) appearing in 
Eq.\eqref{eqn:GF}.}
\end{minipage}
\begin{minipage}{2in}
\begin{tabular}{cr}
\hline
$f^2_{\pi NN}/4\pi$ & 0.08 \\
$f_{\pi N\Delta}$ & 2.206 \\
$f_{\eta NN}$ & 3.889 \\ 
$g_{\rho NN}$ & 8.721 \\ 
$\kappa_\rho$ & 2.654 \\ 
$g_{\omega NN}$ & 8.100 \\ 
$\kappa_\omega$ & 1.020 \\ 
$g^{t}_{\omega NN}$ & 1.298 \\ 
$\kappa^{t}_\omega$ & 1.002 \\ 
$g_{\sigma NN}$ & 6.815 \\ 
$g_{\rho \pi\pi}$ & 4.000 \\ 
$f_{\pi\Delta\Delta}$ & 1.000 \\ 
$f_{\rho N\Delta}$ & 7.516 \\ 
$g_{\sigma\pi\pi}$ & 2.353 \\ 
$g_{\omega\pi\rho}$ & 6.956 \\ 
$g_{\rho\Delta\Delta}$ & 3.302 \\ 
$\kappa_{\rho\Delta\Delta}$ & 2.000 \\
$g_{\rho\pi\gamma}$ & 0.1027$e$\\ 
$g_{\omega\pi\gamma}$ & 0.3247$e$\\ 
$m_\sigma        $&  500.1 MeV \\
\hline
\end{tabular}
\caption{\label{tab:nr_c}Lagrangian bare coupling and $\sigma$ mass.}
\end{minipage}
\begin{minipage}{2in}
\begin{tabular}{cr}
\hline
$\Lambda_{\pi NN}$ & 810 \\ 
$\Lambda_{\pi N\Delta}$ & 829 \\ 
$\Lambda_{\rho NN}$ & 1087 \\ 
$\Lambda_{\rho\pi\pi}$ & 1094 \\ 
$\Lambda_{\omega NN}$ & 1523 \\ 
$\Lambda^{t}_{\omega NN}$ & 589 \\ 
$\Lambda_{\eta NN}$ & 624 \\ 
$\Lambda_{\sigma NN}$ & 781 \\ 
$\Lambda_{\rho N\Delta}$ & 1200 \\ 
$\Lambda_{\pi\Delta\Delta}$ & 600 \\ 
$\Lambda_{\sigma\pi\pi}$ & 1200 \\ 
$\Lambda_{\omega\pi\rho}$ & 600 \\ 
$\Lambda_{\rho\Delta\Delta}$ & 600 \\
\hline
\end{tabular}
\caption{\label{tab:nr_l}Lagrangian non-resonant cutoffs (MeV).}
\end{minipage}
\end{table}

\begingroup
\squeezetable
\begin{table}
\newcolumntype{d}[1]{D{.}{.}{#1}}
\begin{tabular*}{1.025\textwidth}{cc|d{0}d{1}|d{3}|d{3}|d{3}d{3}|d{3}|d{3}d{3}d{3}|d{3}d{3}d{3}|d{3}d{3}}
\# & $L^{(\pi N)}_{TJ}$ & \multicolumn{1}{c}{$M^{(0)}$} &\multicolumn{1}{c}{$k_{N^*}$}& \multicolumn{1}{c}{$\pi N$} & \multicolumn{1}{c}{$\eta N$} 
    & \multicolumn{2}{c}{$\pi \Delta$} & \multicolumn{1}{c}{$\sigma N$} & \multicolumn{3}{c}{$\rho N$} & \multicolumn{3}{c}{$\omega N$}
    &\multicolumn{1}{c}{$A_{\frac{1}{2}}$}&\multicolumn{1}{c}{$A_{\frac{3}{2}}$} \\ 
\hline
1  &$S_{11}$&1800&99.9  &7.049 &9.100 &-1.853&      &-2.795&2.028 &0.027 &      &-3.761&0.405 &      &83.8  &       \\
2  &$S_{11}$&1880&100.0 &9.824 &0.600 &0.045 &      &1.139 &-9.518&-3.014&      &-0.516&0.366 &      &-40.3 &       \\
3  &$S_{31}$&1850&20.7  &5.275 &      &-6.175&      &      &-4.299&5.638 &      &      &      &      &129.4 &       \\
4  &$P_{11}$&1763&76.1  &3.912 &2.621 &-9.905&      &-7.162&-5.157&3.456 &      &-3.362&5.231 &      &-21.8 &       \\
5  &$P_{11}$&2037&22.1  &9.998 &3.661 &-6.952&      &8.629 &-2.955&-0.945&      &-2.095&1.043 &      &-27.5 &       \\
6  &$P_{13}$&1711&76.4  &3.270 &-0.999&-9.988&-5.038&1.015 &-0.003&2.000 &-0.081&5.737 &-0.548&-0.204&-12.4 &-63.8  \\
7  &$P_{31}$&1900&100.0 &6.803 &      &2.118 &      &      &9.915 &0.153 &      &      &      &      &54.1  &       \\
8  &$P_{33}$&1603&83.9  &1.312 &      &1.078 &1.524 &      &2.012 &-1.249&0.379 &      &      &      &-78.6 &-131.2 \\
9  &$P_{33}$&1391&-93.3 &1.319 &      &2.037 &9.538 &      &-0.317&1.036 &0.766 &      &      &      &-6.7  &5.3    \\
10 &$D_{13}$&1899&-35.3 &0.445 &-0.017&-1.950&0.978 &-0.482&1.133 &-0.314&0.179 &-0.081&3.740 &0.230 &88.8  &-71.4  \\
11 &$D_{13}$&1988&-41.7 &0.465 &0.357 &9.919 &3.876 &-5.499&0.289 &9.628 &-0.141&7.883 &9.900 &3.386 &-54.5 &46.8   \\
12 &$D_{15}$&1898&0.0   &0.312 &-0.096&4.792 &0.020 &-0.455&-0.179&1.249 &-0.101&0.625 &1.086 &-0.156&33.0  &40.3   \\
13 &$\bm{D_{15}}$&2334&9.7   &0.167&-0.106 &0.190&-0.098 &-0.075 &-0.530 &0.228&0.099&-0.150 &-1.990 &0.199&12.6 &87.4  \\
14 &$D_{33}$&1976&36.7  &0.945 &      &3.999 &3.997 &      &0.162 &3.949 &-0.856&      &      &      &95.9  &-6.1   \\
15 &$F_{15}$&2187&92.1  &0.062 &0.000 &1.040 &0.005 &1.527 &-1.035&1.607 &-0.026&-0.046&2.212 &0.078 &-99.8 &-68.1  \\
16 &$F_{35}$&2162&-84.2 &0.174 &      &-2.961&-1.093&      &-0.076&8.034 &-0.061&      &      &      &-61.0 &-103.4 \\
17 &$F_{37}$&2137&-100.0&0.254 &      &-0.316&-0.023&      &0.100 &0.100 &0.100 &      &      &      &45.9  &47.7   \\
\end{tabular*}
\caption{\label{tab:resc}Bare masses $M^{(0)}$ (MeV) appearing in the 
resonance propagator of Eq.\eqref{eqn:GF}, and the ranges $k_{N^*}$ (MeV), 
strong couplings $G^{JT}_{LSMB,N^*}$ (MeV$^{-1/2}$) 
and photo-couplings $A^{JT}_{\lambda,p}$ ($10^{-3}$ GeV$^{-1/2}$)
in Eqs.\eqref{eqn:Gmb} and \eqref{eqn:Ggn}.}

\vspace{.2cm}

\newcolumntype{d}[1]{D{.}{.}{#1}}
\begin{tabular*}{0.90\textwidth}{cc|d{3}|d{3}|d{3}d{3}|d{3}|d{3}d{3}d{3}|d{3}d{3}d{3}}
\# & $L^{(\pi N)}_{TJ}$ & \multicolumn{1}{c}{$\pi N$} & \multicolumn{1}{c}{$\eta N$} 
   & \multicolumn{2}{c}{$\pi \Delta$} & \multicolumn{1}{c}{$\sigma N$} & \multicolumn{3}{c}{$\rho N$} & \multicolumn{3}{c}{$\omega N$} \\ 
\hline
1  &$S_{11}$&1676.4&599.0 &554.0 &      &801.0 &1999.9&1893.7&      &500.2 &817.9 &       \\
2  &$S_{11}$&533.5 &500.0 &1999.1&      &1849.5&796.8 &500.0 &      &503.1 &622.0 &       \\
3  &$S_{31}$&2000.0&      &500.0 &      &      &500.0 &500.0 &      &      &      &       \\
4  &$P_{11}$&1203.6&1654.8&729.0 &      &1793.1&622.0 &1698.9&      &675.8 &516.9 &       \\
5  &$P_{11}$&646.9 &897.8 &501.3 &      &1161.2&500.1 &922.3 &      &533.7 &950.1 &       \\
6  &$P_{13}$&1374.0&500.2 &500.0 &500.8 &640.5 &500.0 &500.1 &1645.2&500.1 &547.3 &513.3  \\
7  &$P_{31}$&828.8 &      &2000.0&      &      &1998.8&2000.0&      &      &      &       \\
8  &$P_{33}$&746.2 &      &846.4 &781.0 &      &585.0 &500.2 &1369.7&      &      &       \\
9  &$P_{33}$&880.7 &      &507.3 &501.7 &      &606.8 &1043.4&528.4 &      &      &       \\
10 &$D_{13}$&1658.0&1918.2&976.4 &1034.5&1315.8&599.8 &1615.1&1499.5&565.5 &802.9 &978.1  \\
11 &$D_{13}$&1094.0&678.4 &1960.0&660.0 &1317.0&550.1 &597.6 &1408.7&500.5 &506.2 &545.2  \\
12 &$D_{15}$&1584.7&1554.0&500.8 &820.2 &507.1 &735.4 &749.4 &937.5 &1036.0&999.2 &996.0  \\
13 &$\bm{D_{15}}$&1223.8&1990.2&1910.4&996.1 &921.6 &1022.0&1941.9&997.0 &930.2 &998.2 &999.2  \\
14 &$D_{33}$&806.0 &      &1359.4&608.1 &      &1515.0&1999.0&956.6 &      &      &       \\
15 &$F_{15}$&1641.6&655.9 &1899.5&522.7 &500.9 &500.8 &500.0 &1060.9&541.8 &502.0 &651.6  \\
16 &$F_{35}$&1035.3&      &1228.0&586.8 &      &1514.8&593.8 &1506.0&      &      &       \\
17 &$F_{37}$&1049.0&      &1180.2&1031.8&      &600.0 &600.0 &600.0 &      &      &       \\
\end{tabular*}
\caption{\label{tab:resl}Resonance strong form factor cutoff 
parameters in MeV.}
\end{table}
\endgroup


Large values of \gonn\ have been deduced
from studies of the nucleon electromagnetic form factors \cite{Hohler:1976ax} 
and various $NN$ studies \cite{Nakayama:1998zv},\cite{Machleidt:1987hj}. 
These studies yield a range of $10 \lesssim \gonn \lesssim 20$.
Values of this order were assumed for the studies in Ref.\cite{OTL}
though with a strongly suppressing form factor due to a small cutoff,
$\Lambda_{OTL}=0.5$ GeV. At early stages of the fit when we attempted to use
the values \gonn, \komg, and \Lonn\ determined in fits to the \pnpn\ data
the resulting cross sections were too large
by one or two orders of magnitude for both \pmpon\ and \gpop\ reactions.
In order to reproduce the data within the present model for the limited
parameter search which we have performed it was necessary to introduce
the non-resonant coupling parameters \gonna, \komga, and \Lonna. 
These parameters appear at vertices in the graphs of
Figs.\ref{fig:vonpn}(a),(b), Figs.\ref{fig:vonon} and
Figs.\ref{fig:vgn}(g),(h). The interactions correspond, in a 
four-dimensional formulation to vertices with timelike momentum
transfer. In the fits to the \pnpn\ data $\omega$ mesons appear only
in graphs corresponding to spacelike momentum transfer. An similar
situation obtains in $pp\to pp\pi^0$ reactions
\cite{Koltun:1965yk,*Schillaci:1969hq,*Miller:1991pi,*Lee:1993xh}
where different couplings are used for exchanged and emitted $s$-wave
pions. The small value obtained for the \gonna\ coupling is near
the result found by the Giessen group's study \cite{Shklyar:2004ba}.
The treatment here is certainly phenomenological but no more so
than introducing other non-resonant reaction mechanisms involving
heavy mesons ({\em eg.}\ including \vonrn\ or $f_0/\sigma$ exchange in
\vonpn) or resonances. These alternatives should nevertheless be explored
as a guide, at least, to the model dependencies of the present approach.

We have found that the introduction of the \on\ channel significantly
modifies the behavior of the \pnpn\ $D_{15}$ partial wave amplitude. This
can be seen in Figs.\ref{fig:rtpnpn1} and \ref{fig:itpnpn1} where
we show as a dashed-line curve in the $D_{15}$ panel
the optimal curves found in the first stage fit to \pnpn,
\pmpon, and \gpop\ data described above. 
The $A^{51}_{\thalf}$ photocoupling is large and could be an important
effect in, for example, the electroproduction reaction. Comparison
of the $N^*\to\on$ physical masses and
branching fractions determined in this work with other calculations 
(as in Ref.\cite{PDG}) require the analytic continuation
of the $T$ matrix amplitudes to the physical pole position; this work is
in preparation.


The prediction for the PBA \cite{Titov:1998bw} in \gpop
\begin{align}
\label{eqn:sigb}
\Sigma_\omega(\theta;E) &= 
\frac{\sigma_{\perp}-\sigma_{||}}
     {\sigma_{\perp}+\sigma_{||}}
\end{align}
is shown in Fig.\ref{fig:gosig}. 
Here $\sigma_{||}(\sigma_{\perp})$ is the differential cross section
for linearly polarized photons in (perpendicular to) the emission plane
of the $\omega$ meson. At the lowest energy $E=1.743$
GeV a study is made of the sensitivity to the resonance contribution for
three cases. The thin-dashed line is the result when all the resonance
contributions have been removed. Other curves in the figure show
the result when one of three dominate waves is removed.

\begin{figure}[b]
\includegraphics[ width=300 pt, keepaspectratio, clip]{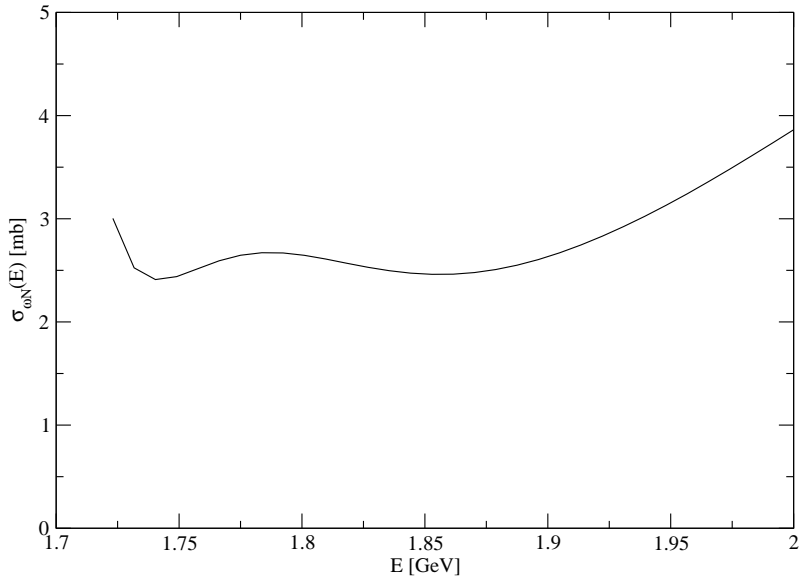}
\caption{\label{fig:sig_onon}Calculated total elastic \on\ cross section
(mb) as a function of center-of-mass energy $E=W$.}
\end{figure}

The total cross section of the reaction \onon\ is of interest
for realistic calculations of nuclear matter properties. The predicted total
cross section for this reaction is shown in Fig.\ref{fig:sig_onon}.
The scattering lengths obtained from the $T$ matrix are
\begin{align}
a_{J} &= \lim_{E\to m_\omega+m_N} 
\frac{\pi m_\omega m_N}{m_\omega+m_N}T^J_{0J\on,0J\on}(E), \\
a_{\half} &= [-0.0454 - i0.0695]\mbox{ fm}, \\
a_{\thalf} &= [0.180 - i 0.0597]\mbox{ fm},
\end{align}
related to the total cross section at threshold by
$\sigma_{\on}(E\to m_\omega+m_N) = 4\pi(|a_{\half}|^2+2|a_{\thalf}|^2)/3$.

\begin{figure}[t]
\includegraphics[ width=300 pt, keepaspectratio, clip]{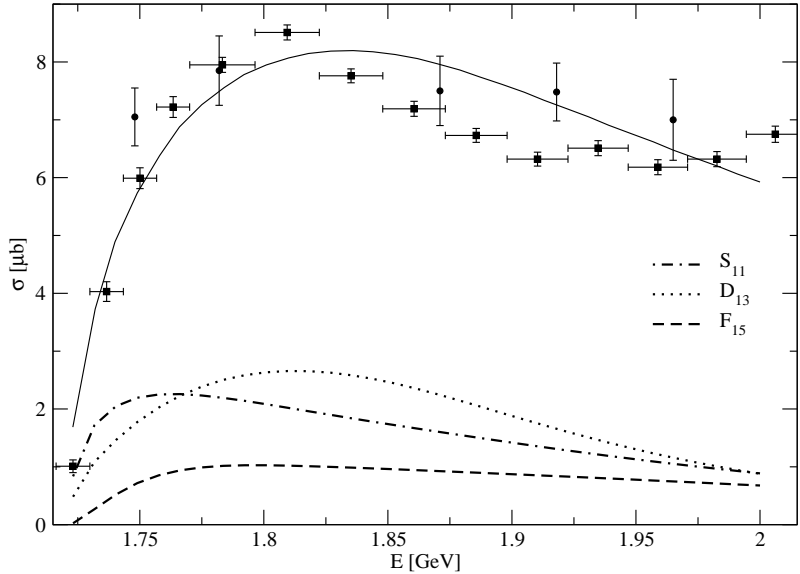}
\caption{\label{fig:sig_gpop}Total cross section for \gpop\ ($\mu$b)
compared with that extracted from data from Ref.\cite{Klein:1998xy} 
(circles) and Ref.\cite{Barth:2003kv} (squares) as a function of 
center-of-mass energy $E=W$. The error bars on the data from 
Ref.\cite{Barth:2003kv} are statistical errors only. Systematic errors
are about $10-15\%$ \cite{Barth:2003kv}. The three partial waves with 
the largest contribution are shown. See text for discussion.}
\end{figure}

\begin{figure}[ht]
\includegraphics[ width=350 pt, keepaspectratio, clip]{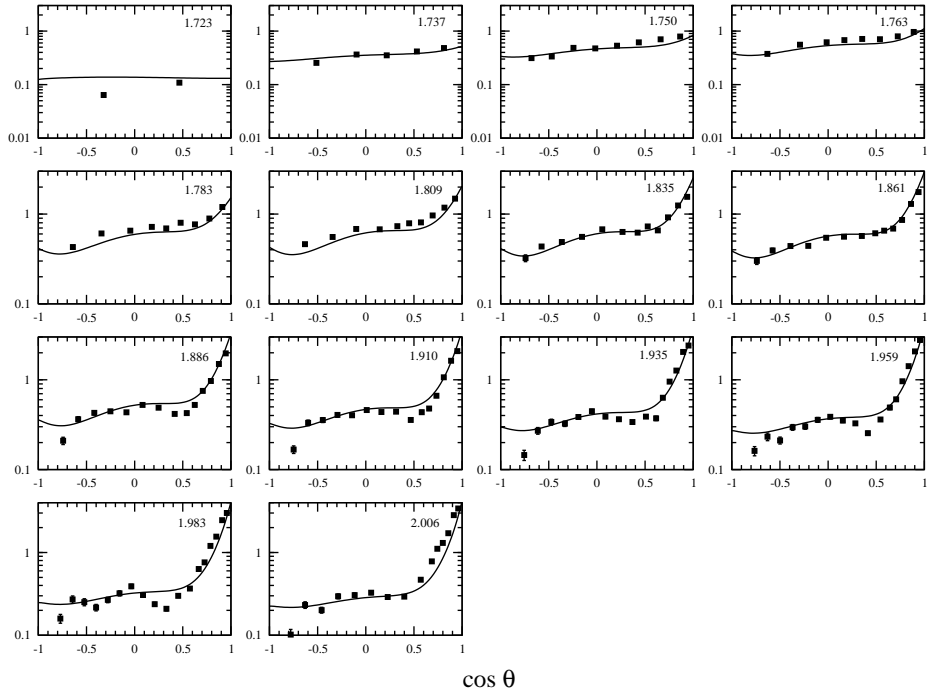}
\caption{\label{fig:godxsx}Semi-logarithmic version of Fig.\ref{fig:godxs},
unpolarized DCS for \gpop\ ($\mu$b/sr) vs. $\cos\theta$. 
Center-of-mass energies are shown in the upper-right corner of each 
panel. The sharp angular features near are accenuated $x\simeq 0.5$ 
for $E>1.86$ GeV and $x\simeq -0.5$ for $E\ge 1.935$.}
\end{figure}

The total cross section for \gpop\ is shown in Fig.\ref{fig:sig_gpop}
along with contributions from partial waves $\ell_{TJ}$ with significant
contributions. The error bars on the total cross section are 
statistical\cite{Barth:2003kv}. Systematic errors are shown in 
Ref.\cite{Barth:2003kv} to be about $10-15\%$. They 
arise from, among other sources, the
extrapolation of the DCS in the forward and backward directions 
for center-of-mass $\omega$-meson scattering angles $\theta < 15^\circ$
and $\theta > 150^\circ$. The systematic errors for the DCS from 
Ref.\cite{Barth:2003kv} are largest in the backward direction where the
discrepancy from our calculated cross section, as seen in 
Fig.\ref{fig:godxsx} is most pronounced. Nevertheless, the present 
calculated \gpop\ DCS appears to miss some small angle structure near
$x\simeq 0.5$ above $E>1.86$ GeV and $x\simeq -0.5$ and
0.5 for $E\ge 1.935$ GeV,
possibly a result of destructive interference effects of some combination 
of additional higher mass resonances and non-resonant effects not included 
in this study.\footnote{This feature is also seen in recent precision data
from the CLAS Collaboration\cite{Williams:2007phd}.}

\section{Conclusion\label{sec:conclusion}}
A dynamical coupled channel model for six-channels has been employed
in simultaneous fits of the pion and photon induced single pion and
omega production reactions.  The \pn\ partial wave amplitudes,
unpolarized differential cross sections
(DCS) and photon beam asymmetry (PBA) have been fit with $\chi^2\sim 1$
for center-of-mass energies from threshold to $E<1.65$ GeV. At higher
energies, the model is unable to accurately
reproduce the data. There are several
ways one might attempt to remedy this deficiency. If we work within
the present model formulation and keep the same non-resonant mechanisms
it is possible that a more thorough search of the parameters may yield a
better fit at higher energies. The introduction of more resonances may
also yield a better fit. However, for energies $W>1.8$ GeV the existing 
data makes distinguishing non-resonant mechanisms from resonant 
mechanisms difficult. Particularly useful in this endeavor would be 
more high precision single and some double
polarization observables for both pion and $\omega$-meson production.

Additional mechanisms in the non-resonant terms are sure to contribute,
perhaps significantly, to the calculated scattering observables at these
higher energies. At the two-body level we have neglected couplings such
as \vonen, \vonpd, \vonsn, and \vonrn. There may also be significant effects
from additional mechanisms in the \vongn\ interaction. We have neglected
the effects of $t$-channel $\sigma$ exchange, $\eta$ exchange (generally
thought to be small) and Pomeron exchange in \gpop, known to have large 
contributions at forward angles at high energies. At energies above
the two-pion production threshold, the $\pi\pi N$ channel contribution
can give a significant contribution and must be calculated. This can
be accomplished in the present model formulation and is currently under
study.

A fair prediction for the $\omega$ meson PBA, $\Sigma(\theta;E)$
has been obtained near threshold, shown in the upper-left panel 
in Fig.\ref{fig:gosig}. At higher energies, $W \gtrsim 1.8$ GeV the 
calculated beam asymmetry does not agree well with the shape of the
data. This observable is sensitive to both resonant and non-resonant
contributions could improve with any of the refinements discussed
above.

The present model will be used to analyze the
electroproduction data in the region $Q^2 \lesssim 5$ GeV$^2$ and the
extension to the photoproduction of $\rho$ and $\phi$ vector mesons.
Photoproduction of $\omega$ from nuclei gives information about the
spectral function of the $\omega$ meson in the nuclear 
medium\cite{Muhlich:2003tj}. Inclusion of off-shell effects in such
an analysis is required and the present model affords a starting point
for their inclusion.

\appendix*
\section{$P_{33}(1232)$ transition form factor}
The transition form factor for $\gp\to\Delta(1232)$ is taken 
as\cite{Jones:1972ky}
\begin{align}
\Gamma^{33}_{\thalf,p}(q) &= -\mathcal{K}_M G_M(q^2)
 +(\mathcal{K}_{M}+\mathcal{K}_{E})G_E(q^2) \\
\Gamma^{33}_{\half,p}(q) &= \frac{1}{\sqrt 3} [-\mathcal{K}_M G_M(q^2)
 +(\mathcal{K}_{M}-\mathcal{K}_{E})G_E(q^2)] \\
\mathcal{K}_M &= \frac{e}{(2\pi)^{3/2}}
\sqrt{\frac{E_N(q)+m_N}{2E_N(q)}}\frac{1}{\sqrt{2|q^0|}}
\frac{3(m_\Delta+m_N)}{2m_N}\frac{E\qmag\sqrt\pi}{Q^2+(m_\Delta+m_N)^2} \\
\mathcal{K}_{E} &= -\frac{4E\qmag^2}{E_N(q)+m_N}
   \frac{1}{Q^2+(m_\Delta-m_N)^2}\mathcal{K}_M.
\end{align}
On resonance at the photon point, $Q^2=0$ the $G_M(0)$ and $G_E(0)$ are
related to the value $A^{33}_{\lambda,p}$ in Table \ref{tab:resc} as
\begin{align}
A^{33}_{\thalf,p} &=-\frac{\sqrt{3}}{2}\frac{e}{2m_N}
\sqrt{\frac{m_\Delta\qmag}{m_N}} [G_M(0)+G_E(0)] \\
A^{33}_{\half,p} &=-\frac{1}{2}\frac{e}{2m_N}
\sqrt{\frac{m_\Delta\qmag}{m_N}} [G_M(0)-3G_E(0)].
\end{align}
The values in the table correspond to
\begin{align}
G_M(0) &= 1.62  \\
G_E(0) &= 0.015.
\end{align}

\begin{acknowledgments}
The author wishes to thank T.\ Sato for Born amplitudes, 
F.\ Klein and M.\ Williams for providing data, and to 
T.-S.H.\ Lee and A.W.\ Thomas for useful discussions.
This work is supported by the U.S.\ Department of Energy, Office of
Nuclear Physics Division under contract No. DE-AC02-06CH11357, and
contract No. DE-AC05-060R23177 under which Jefferson Science Associates
operates Jefferson Lab.
This research used resources of the National Energy Research Scientific 
Computing Center, which is supported by the Office of Science of the 
U.S. Department of Energy under Contract No. DE-AC02-05CH11231.
\end{acknowledgments}

\mciteErrorOnUnknownfalse
\bibliography{master}

\end{document}